\documentstyle[titlepage,12pt]{article}

\begin{document}
\title{Weiss-approach to pair of coupled non-linear reaction-diffusion
equations}
\vspace{5in}
\author{A.L.Larsen\thanks{E-mail: ALLARSEN@nbivax.nbi.dk}\\Nordita,
Blegdamsvej 17, Dk-2100 Copenhagen \O, Denmark}
\maketitle
\begin{abstract}
We consider a pair of coupled non-linear partial differential equations
describing a biochemical model system. The Weiss-algorithm for the
Painlev\'{e} test, that has been succesfully used in mathemathical
physics for the KdV-equation, Burgers equation, the sine-Gordon equation etc.,
is applied, and we find
that the system possesses only the "conditional" Painlev\'{e} property. We
use the outcome of the analysis to construct an auto-B\"{a}cklund
transformation, and we find a variety of one and two-parameter families
of special solutions.
\end{abstract}
\section{Introduction}
Although the deeper reasons are still not completely understood, it is clear
today that the Painlev\'{e} property [1] of (systems of) ordinary or partial
differential equations is closely related to the concept of integrability.
The original approach of Ablowitz, Ramani and Segur [2], led to the conjecture
that a non-linear partial differential equation (PDE) is integrable if every
exact reduction to an ordinary differential equation (ODE) has the Painlev\'{e}
property, i.e. has no other movable singularities than poles. This version of
the Painlev\'{e} conjecture is well-suited to confirm already known
integrability properties of a PDE, but is less helpfull when it comes to the
act
of finding new integrable systems, due to the problems of actually finding all
the exact reductions. In this respect the later approach of Weiss, Tabor and
Carnevale [3] seems to be much more usefull. In the simplest version of this
approach a PDE is conjectured to be integrable if its solutions are
singlevalued about movable "singularity-manifolds":
\begin{equation}
\phi(z_1,z_2,...,z_n)=0,
\end{equation}
where $\phi$ is an arbitrary ("movable") analytic function. In other words a
solution $x(z_i)$ to the PDE in question should have a Laurent-like expansion
about the movable singular manifold (1.1):
\begin{equation}
x(z_i)=[\phi(z_i)]^\rho\sum_{j=0}^{\infty}[\alpha_j(z_i)][\phi(z_i)]^j,
\end{equation}
where $\rho$ is a negative integer. If the number of arbitrary functions in
the expansion (1.2) equals the order of the PDE, we have reason to believe
that (1.2) is the generic expansion about $\phi=0$. If this is not the case
we may have lost something and the system does not have the full Painlev\'{e}
property. Further details can be found in Refs. 1,3.

This Weiss-approach to the Painlev\'{e} property has been used to analyse a
long list of the PDE's of mathematical physics like the KdV-equation, Burgers
equation, sine-Gordon equation, modified KdV-equation, Boussinesq equation,...
[3,4]. The method is of course also applicable to
systems of coupled non-linear PDE's like the coupled KdV-equations
(Hirota-Satsuma system) [5]:
\begin{eqnarray}
x_t\hspace*{-2mm}&=&\hspace*{-2mm}\frac{1}{2}x_{rrr}+3xx_r-6yy_r,\nonumber\\
y_t\hspace*{-2mm}&=&\hspace{-2mm}-y_{rrr}-3xy_r,\nonumber
\end{eqnarray}
or the modified sine-Gordon system [5]:
\begin{eqnarray}
x_{t}+\frac{1}{2}xy+\frac{\lambda}{2}\frac{x}{y}
\hspace*{-2mm}&=&\hspace*{-2mm}0,\nonumber\\
y_{r}+\frac{1}{2}xy+\frac{1}{2\lambda}\frac{y}{x}
\hspace*{-2mm}&=&\hspace*{-2mm}0.\nonumber
\end{eqnarray}

Systems of coupled non-linear PDE's also arise naturally in (mathematical)
biology and chemistry, for instance in models describing reactions and
diffusion of different components in a medium. In this paper we will
consider the following system of PDE's [6-10]:
\begin{eqnarray}
x_t\hspace*{-2mm}&=&\hspace*{-2mm}x^2y-Bx+Kx_{rr},\nonumber\\
y_t\hspace*{-2mm}&=&\hspace*{-2mm}-x^2y+Bx+Ky_{rr},
\end{eqnarray}
where $B$ and $K$ are constants. These equations are supposed to describe the
evolution of concentrations $x$ and $y$ of 2 chemical components.
Obviously $K$ is a
diffusion coefficient and the terms $x^2y-Bx$ represent the irreversible
reactions. Such processes are certainly not expected to be integrable and we
will indeed see that (1.3) does not have the full Painlev\'{e} property
(Sec. 2). We note in passing that the system is closely related to the
so-called Brusselator [6] describing the scheme of reactions:
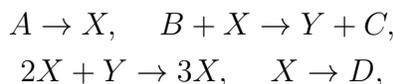
\begin{eqnarray}
&A\rightarrow X,\hspace*{5mm}B+X\rightarrow Y+C,&\nonumber\\
&2X+Y\rightarrow 3X,\hspace*{5mm}X\rightarrow D,&\nonumber
\end{eqnarray}
where $A,B,C,D,X$ and $Y$ are different kinds of molecules.

The system of equations (1.3) has been investigated by several
authors [6-10]. The steady state solutions $(x_t=y_t=0)$ were analysed in Ref.
7 under the boundary condition $x+y=const.$ In that case Eqs. (1.3) trivially
separate and are completely solvable in terms of elliptic functions. In
Refs. 8 and 9 a system like (1.3), but with 2 different diffusion coefficients,
was investigated. It was assumed that one of the components diffuses very
rapidly so that one of the diffusion coefficients could be taken to be
infinite.
The system was then reduced to a one-variable non-linear PDE and some
travelling wave solutions were found. Finally we mention that a one-parameter
family of travelling wave solutions for the coupled system (1.3) was found in
Ref. 8 under the boundary condition $x+y=const.$

The paper is organized as follows: In section 2 we carry out the Painlev\'{e}
analysis of the system (1.3) using the Weiss-approach. We consider expansions
of $x$ and $y$ about a movable singular manifold and find that the system
possesses only the so-called conditional Painlev\'{e} property. In section 3
we truncate the expansions of section 2 at the most singular term, which allows
us to construct a two-parameter family of special solutions in terms of
trigonometric functions. In section 4 we truncate the expansions of section 2
at the "constant" term. This leads to an auto-B\"{a}cklund transformation
between 2 pairs of solutions. We then construct some more one and
two-parameter families of special solutions, and in section 5 we finally give
our conclusions and we outline how
more general families of solutions can be "chased".
\newpage
\section{Painlev\'{e} test}
\setcounter{equation}{0}
In this section we perform the Painlev\'{e} analysis of the system of coupled
non-linear PDE's (1.3) with $B$ and $K$ as arbitrary positive constants (the
positivity of $B$ and $K$ is of course not mandatory, at least not from a
mathematical point of view).

The first step in the Weiss-algorithm [3] is to look for the dominant behaviour
about a movable singular manifold $\phi(r,t)=0$. Thus we write:
\begin{equation}
x=U_o(r,t)[\phi(r,t)]^\rho\equiv U_o\phi^\rho;\hspace*{5mm}\rho<0,
\end{equation}
\begin{equation}
y=V_o(r,t)[\phi(r,t)]^\sigma\equiv V_o\phi^\sigma;\hspace*{5mm}\sigma<0,
\end{equation}
and balances the most singular terms after insertion in (1.3). This leads to
the unique solution:
\begin{equation}
\rho=\sigma=-1,\hspace*{5mm}U_o=-V_o,
\end{equation}
with:
\begin{equation}
U_o=\pm\sqrt{2K}\phi_r.
\end{equation}
In the following we only consider the case of $+$-sign in (2.4). The other
possibility corresponds to an overall change of sign of $x$ and $y$. Clearly
if $(x,y)$ solves (1.3) also $(-x,-y)$ is a solution.

The next step is to look for the "resonances" [3], i.e. the orders in the
expansions where arbitrary functions may appear. Keeping (2.3) and (2.4) in
mind we write:
\begin{equation}
x=\sqrt{2K}\phi_r\phi^{-1}+s_1\phi^{l-1},
\end{equation}
\begin{equation}
y=-\sqrt{2K}\phi_r\phi^{-1}+s_2\phi^{l-1},
\end{equation}
where $l$ is a non-negative integer and $s_i=s_i(r,t); i=1,2$. After insertion
in (1.3) and balancing of the most singular terms we get the matrix-equation:
\begin{equation}
\left(\begin{array}{cc} (l-1)(l-2)-4 & 2\\
4 & (l-1)(l-2)-2\end{array}\right) \left(\begin{array}{c}s_1\\s_2\end{array}
\right)
\equiv{\cal A}\left(\begin{array}{c}s_1\\s_2\end{array}\right)=\left(\begin
{array}{c}
0\\0\end{array}\right),
\end{equation}
with:
\begin{equation}
\det{\cal A}=(l-1)(l-2)[(l-1)(l-2)-6].
\end{equation}
For a given root of the polynomium (2.8) there are now as many arbitrary
functions $s_i$ as the multiplicity of that particular root. Solving the
equation $\det{\cal A}=0$ we find the roots $(-1,1,2,4)$. The root $l=-1$
corresponds to the arbitraryness of the location of the singular manifold
$\phi$, while the roots $l=1$, $l=2$ and $l=4$ are supposed to correspond to
arbitrary functions at the orders $\phi^0$, $\phi^1$ and $\phi^3$ in the
expansions of $x$ and $y$ about $\phi=0$. Note that the number of roots equals
the order of the system (1.3) so until now everything works
fine.

The third and final step in the Weiss-algorithm [3] is to expand out to the
highest resonance to make sure that no inconsistencies arise, i.e. we write:
\begin{equation}
x=\sqrt{2K}\phi_r\phi^{-1}+\sum_{j=0}^3\alpha_j\phi^j;\hspace*{5mm}
\alpha_j=\alpha_j(r,t),
\end{equation}
\begin{equation}
y=-\sqrt{2K}\phi_r\phi^{-1}+\sum_{j=0}^3\beta_j\phi^j;\hspace*{5mm}
\beta_j=\beta_j(r,t).
\end{equation}
These expressions are inserted into (1.3) and we then balance the terms order
by order in $\phi$. It was demonstrated by Kruskal [11] that some
simplifications arise in this process if one formally solves the equation
$\phi(r,t)=0$ for (say) $r$ and then writes:
\begin{equation}
\phi(r,t)=r+\psi(t),\hspace*{5mm}\alpha_j=\alpha_j(t),\hspace*{5mm}
\beta_j=\beta_j(t),
\end{equation}
where $\psi$ is an arbitrary function. For our purposes of constructing
explicit special solutions (sections 3,4) it is however necessary to use the
general expansions (2.9) and (2.10).

At the various orders we now get:

\hspace*{-6mm}$\phi^{-2}$:
\begin{equation}
\sqrt{2K}(2\alpha_0-\beta_0)=\frac{\phi_t}{\phi_r}-3K\frac{\phi_{rr}}{\phi_r},
\end{equation}
so here we get an arbitrary function (say) $\alpha_0$, as expected from the
analysis of the resonances.

\hspace*{-6mm}$\phi^{-1}$:
\begin{equation}
\sqrt{2K}(2\alpha_1-\beta_1)=\frac{\alpha_0 (2\beta_0-\alpha_0)}{\phi_r}-
\frac{\phi_{rt}}{\phi_r^2}-\frac{B}{\phi_r}+K\frac{\phi_{rrr}}{\phi_r^2},
\end{equation}
with another arbitrary function (say) $\alpha_1$.

\hspace*{-6mm}$\phi^0$:
\begin{equation}
\left\{\begin{array}{ll}\alpha_2=F_1(\alpha_0,\alpha_1,\phi;B,K)\\
\beta_2=F_2(\alpha_0,\alpha_1,\phi;B,K)\end{array}\right. ,
\end{equation}
where $F_1$ and $F_2$ are certain complicated expressions in the arbitrary
functions $\phi$, $\alpha_0$, $\alpha_1$ and their derivatives, as well as in
the "chemical" constants $B$ and $K$. For convenience they are listed in the
appendix.

\hspace*{-6mm}$\phi^1$:
\begin{equation}
\left\{\begin{array}{ll}\alpha_3+\beta_3=G_1(\alpha_0,\alpha_1,\phi;B,K)\\
\alpha_3+\beta_3=G_2(\alpha_0,\alpha_1,\phi;B,K)\end{array}\right. .
\end{equation}
There is now an arbitrary function at this order if and only if the 2 right
hand sides are equal. From the explicit expressions for $G_1$ and $G_2$ given
in the appendix it follows that this is not so (this is actually most easily
seen by using the Kruskal {\it Ansatz} (2.11) and by keeping in mind the
arbitraryness of (say) $\alpha_0$ and $\alpha_1$). It follows that (say)
$\alpha_3$ is arbitrary only if the singularity manifold $\phi$ satisfies a
certain constraint. Therefore, the system (1.3) does not have the full
Painlev\'{e} property but only the "conditional" one [12].
\section{Truncation and special solutions}
\setcounter{equation}{0}
In this and the following section we will extensively use the expressions
among the alpha's and beta's obtained in section 2. In this section we look
for special solutions to (1.3) obtained by truncation of the expansions (2.9)
and (2.10) at the singular terms. Taking $\alpha_i=\beta_i=0; i\geq0$ we find:
\begin{equation}
\left(\begin{array}{c}x\\y\end{array}\right)=\sqrt{2K}\frac{\phi_r}{\phi}
\left(\begin{array}{c}1\\-1\end{array}\right).
\end{equation}
This can however only be a solution to (1.3)
provided equations (2.12)-(2.15) are fulfilled.
It is easily seen that quations (2.14) and (2.15) are trivially
satisfied whereas (2.12) and (2.13) give the compatibility conditions:
\begin{equation}
\frac{\phi_t}{\phi_r}-3K\frac{\phi_{rr}}{\phi_r}=0,
\end{equation}
\begin{equation}
\frac{B}{\phi_r}+\frac{\phi_{rt}}{\phi_r^2}-K\frac{\phi_{rrr}}{\phi_r^2}=0.
\end{equation}
These 2 equations are integrated and consistently solved by:
\begin{equation}
\phi(r,t)=e^{-\frac{3}{2}Bt}\left( c_1\cos(\sqrt{\frac{B}{2K}}r)+
c_2\sin(\sqrt{\frac{B}{2K}}r)\right)+c_3,
\end{equation}
where $c_1,c_2,c_3$ are arbitrary constants. From (3.1) we then get:
\begin{equation}
x(r,t)=-y(r,t)=\frac{\sqrt{B}e^{-\frac{3}{2}Bt}\left( L_1\cos(\sqrt{\frac
{B}{2K}}r)-L_2\sin(\sqrt{\frac{B}{2K}}r)\right)}{e^{-\frac{3}{2}Bt}
\left(L_2\cos(\sqrt{\frac{B}{2K}}r)+L_1\sin(\sqrt{\frac{B}{2K}}r)\right)+1},
\end{equation}
representing a two-parameter family ($L_1$ and $L_2$ being 2 arbitrary
constants to be determined by the initial/boundary conditions) of solutions
to the system (1.3).
\section{Auto-B\"{a}cklund transformation and special solutions}
\setcounter{equation}{0}
In this section we truncate the expansions (2.9) and (2.10) at order $\phi^0$,
i.e. we take $\alpha_i=\beta_i=0; i\geq 1$, and look for solutions in the form:
\begin{equation}
\left(\begin{array}{c}x\\y\end{array}\right)=\sqrt{2K}\frac{\phi_r}{\phi}
\left(\begin{array}{c}1\\-1\end{array}\right)+\left(\begin{array}{c}
\alpha_0\\ \beta_0\end{array}\right).
\end{equation}
As in section 3 this can of course only be a solution to (1.3) provided
(2.12)-(2.15) are fulfilled. Equations (2.12) and (2.13) read:
\begin{equation}
\sqrt{2K}(2\alpha_0-\beta_0)\phi_r=\phi_t-3K\phi_{rr},
\end{equation}
\begin{equation}
\alpha_0(2\beta_0-\alpha_0)=\frac{\phi_{rt}}{\phi_r}+B-K\frac{\phi_{rrr}}
{\phi_r}.
\end{equation}
After using (6.1) and (6.2) from the appendix, equation (2.14) leads to:
\begin{eqnarray}
\alpha_{0t}\hspace*{-2mm}&=&\hspace*{-2mm}\alpha_0^2\beta_0-B\alpha_0+
K\alpha_{0rr},\nonumber\\
\beta_{0t}\hspace*{-2mm}&=&\hspace*{-2mm}-\alpha_0^2\beta_0+B\alpha_0+
K\beta_{0rr},
\end{eqnarray}
while (2.15) is trivially fulfilled. (4.2)-(4.4) represents an overdetermined
system of equations to be solved for $\alpha_0, \beta_0$ and $\phi$. This
system is extremely complicated in the general case but fortunately it is not
so difficult to find special solutions.  Note also that (4.4) has the same
form as the original system (1.3), so if we can find solutions ($\alpha_0,
\beta_0, \phi$) to (4.2)-(4.4) it follows that (4.1) is an auto-B\"{a}cklund
transformation between the 2 pairs of solutions $(x,y)$ and
$(\alpha_0,\beta_0)$
to equation (1.3). It means that if we find one pair of solutions we can in
principle always generate new ones.

Special solutions to the overdetermined system (4.2)-(4.4) can conveniently
be parametrized by the 2 functions [12]:
\begin{equation}
C\equiv\frac{\phi_t}{\phi_r},\hspace*{5mm}V\equiv\frac{\phi_{rr}}{\phi_r}.
\end{equation}
For simplicity we will now restrict ourselves by considering only constant $C$
and $V$. It turns out that two different types of solutions are possible
corresponding to $V=0$ and $V\neq 0$. In the case that $V=0$ and $C\equiv C_0$
is an arbitrary constant we find:
\begin{equation}
\phi(r,t)=c_1(r+C_0t)+c_2,
\end{equation}
where $c_1, c_2$ are arbitrary constants. Equations (4.2)-(4.4) lead to:
\begin{eqnarray}
&2\alpha_0-\beta_0=\frac{C_0}{\sqrt{2K}},&\nonumber\\
&\alpha_0(2\beta_0-\alpha_0)=B,&\\
&\alpha_0^2\beta_0-B\alpha_0=0,&\nonumber
\end{eqnarray}
that are solved by:
\begin{equation}
\alpha_0=\beta_0=\pm\sqrt{B},\hspace*{5mm}C_0=\pm\sqrt{2KB}.
\end{equation}
Using (4.1) we then get:
\begin{equation}
x(r,t)=-y(r,t)\pm2\sqrt{B}=\frac{\sqrt{2K}}{(r\pm\sqrt{2KB}t)+L_1}\pm\sqrt{B},
\end{equation}
representing a one-parameter family ($L_1$ being an arbitrary constant) of
solutions to (1.3).

In the case that both $V\equiv V_0$ and $C\equiv C_0$ are arbitrary constants
($V_0\neq 0$) we find instead:
\begin{equation}
\phi(r,t)=c_1 e^{V_0(r+C_0t)}+c_2,
\end{equation}
where $c_1, c_2$ are arbitrary constants. In this case equations (4.2)-(4.4)
lead to:
\begin{eqnarray}
&\sqrt{2K}(2\alpha_0-\beta_0)=C_0-3KV_0,&\nonumber\\
&\alpha_0(2\beta_0-\alpha_0)=V_0C_0+B-KV_0^2,&\\
&\alpha_0^2\beta_0-B\alpha_0=0,&\nonumber
\end{eqnarray}
Solving (4.11) for $(\alpha_0, \beta_0, C_0)$ in terms of $(B, K, V_0)$ leads
to the 3 possibilities:
\begin{equation}
(\alpha_0, \beta_0, C_0)=\left(-\sqrt{2K}V_0, \frac{-B}{\sqrt{2K}V_0},
\frac{B}{V_0}-KV_0\right),
\end{equation}
and:
\begin{equation}
(\alpha_0, \beta_0, C_0)=\left(\frac{-\sqrt{2K}V_0\pm\sqrt{W_0}}{2}, \frac{2B}{
-\sqrt{2K}V_0\pm\sqrt{W_0}}, \frac{W_0\sqrt{2K}\mp KV_0\sqrt{W_0}}{
-\sqrt{2K}V_0\pm\sqrt{W_0}}\right),
\end{equation}
where $W_0\equiv 2B+KV_0^2$. From (4.1) we finally get:
\begin{equation}
x(r,t)=-y(r,t)+\alpha_0+\beta_0=\frac{\sqrt{2K}V_0e^{V_0(r+C_0t)}}{
e^{V_0(r+C_0t)}+L_1}+\alpha_0,
\end{equation}
with $(\alpha_0, \beta_0, C_0)$ given by one of (4.12), (4.13). This equation
finally represents a two-parameter family ($L_1$ and $V_0$ being arbitrary
constants) of solutions to (1.3).
\section{Conclusion}
\setcounter{equation}{0}
In conclusion we have studied the pair of reaction-diffusion equations (1.3)
using the Weiss-algorithm for the Painlev\'{e} test. The system was found to
possess only the conditional Painlev\'{e} property. The results of the
analysis however led to various kinds of special solutions obtained via
truncations and auto-B\"{a}cklund transformations.

The special solutions we have constructed (3.5), (4.9) and (4.14) are all of
the form $x(r,t)+y(r,t)=const.$ Note that addition of the two equations in
(3.1)
leads to:
\begin{equation}
(x+y)_t=K(x+y)_{rr},
\end{equation}
which is just the ordinary one-variable diffusion equation with the well-known
general solution (see for instance reference 13):
\begin{equation}
(x+y)(r,t)=\frac{1}{2\sqrt{\pi
K}}\int_{-\infty}^{\infty}\frac{F(\xi)}{\sqrt{t}}
e^{-\frac{(r-\xi)^2}{4Kt}}d\xi,
\end{equation}
where:
\begin{equation}
(x+y)(r,0)=F(r).
\end{equation}
All our solutions therefore correspond to the boundary condition $F(r)=const.$,
and could in principle have been obtained by applying the Weiss-algorithm to a
one-variable non-linear reaction-diffusion equation obtained from (3.1) by
separating $x$ and $y=-x+const.$ from the beginning.

The advantage of our more general approach where (in the notation of (5.2))
$F(r)$ is not fixed from the beginning , is of course that we can now use the
results of sections 2-4 to look for more general families of solutions with
non-constant $F(r)$: We can either return to equation (4.5) and look for
solutions to (4.2)-(4.4) with non-constant $V$ and/or $C$, or we can continue
along the road of sections 2 and 3 and look at expansions truncated at higher
orders in $\phi$. That is however out of the scope of this paper.
\vskip 12pt
{\bf Acknowledgements}:
I would like to thank J.B.Pedersen for a discussion on general aspects of
diffusion and reactions.
\newpage
\section{Appendix}
\setcounter{equation}{0}
In this appendix we list the functions $(F_1,F_2)$ appearing in (2.14) and the
functions $(G_1,G_2)$ appearing in (2.15).

The functions $F_1$ and $F_2$ are given by:
\begin{eqnarray}
4K\phi_r^2F_1\hspace*{-2mm}&=&\hspace*{-2mm}\beta_{0t}-\beta_1\phi_t+
\alpha_0^2\beta_0-B\alpha_0-K\beta_{0rr}-2K\beta_{1r}\phi_r\nonumber\\
&-&\hspace*{-2mm}K\beta_1\phi_{rr}+2\sqrt{2K}\phi_r(\alpha_0\beta_1+
\alpha_1\beta_0-\alpha_0\alpha_1),
\end{eqnarray}
\begin{eqnarray}
4K\phi_r^2F_2\hspace*{-2mm}&=&\hspace*{-2mm}(2\alpha_0+\beta_0)_t+
(2\alpha_1+\beta_1)\phi_t-K(2\alpha_0+\beta_0)_{rr}\nonumber\\
&-&\hspace*{-2mm}2K(2\alpha_1+\beta_1)_r\phi_r
+B\alpha_0-\alpha_0^2\beta_0-K(2\alpha_1+\beta_1)\phi_{rr}\nonumber\\
&-&\hspace*{-2mm}2K\sqrt{2K}\phi_r(\alpha_0\beta_1+
\alpha_1\beta_0-\alpha_0\alpha_1).
\end{eqnarray}
Using (2.12) and (2.13) it is straightforward to express $F_1$ and $F_2$ in
terms of (say) the arbitrary functions $(\alpha_0, \alpha_1, \phi)$ and their
derivatives as well as the "chemical" constants $(B,K)$.

The functions $G_1$ and $G_2$ are given by:
\begin{eqnarray}
4K\phi_r^2G_1\hspace*{-2mm}&=&\hspace*{-2mm}2\alpha_{1t}+4\alpha_2\phi_t+
2\sqrt{2K}\phi_r(2\alpha_0\alpha_2+\alpha_1^2)+2B\alpha_1\nonumber\\
&-&\hspace*{-2mm}4\beta_0(\sqrt{2K}\phi_r\alpha_2+\alpha_0\alpha_1)
-2\beta_1(2\sqrt{2K}\phi_r\alpha_1+\alpha_0^2)\nonumber\\
&-&\hspace*{-2mm}4\sqrt{2K}\beta_2\phi_r\alpha_0-2K(\alpha_{1rr}+
4\alpha_{2r}\phi_r+2\alpha_2\phi_{rr}),
\end{eqnarray}
\begin{eqnarray}
4K\phi_r^2G_2\hspace*{-2mm}&=&\hspace*{-2mm}\beta_{1t}+2\beta_2\phi_t
-\sqrt{2K}\phi_r(2\alpha_0\alpha_2+\alpha_1^2)-B\alpha_1\nonumber\\
&+&\hspace*{-2mm}2\beta_0(\sqrt{2K}\phi_r\alpha_2+\alpha_0\alpha_1)+
\beta_1(2\sqrt{2K}\phi_r\alpha_1+\alpha_0^2)\nonumber\\
&+&\hspace*{-2mm}2\sqrt{2K}\beta_2\phi_r\alpha_0-K(\beta_{1rr}+
4\beta_{2r}\phi_r+2\beta_2\phi_{rr}).
\end{eqnarray}
Using (6.1), (6.2) and (2.12), (2.13) we can again express the right hand sides
in terms of (say) the arbitrary functions ($\alpha_0, \alpha_1, \phi)$ and
their derivatives as well as the "chemical" constants $(B,K)$. It can then be
verified that $G_1\neq G_2$, so that both $\alpha_3$ and $\beta_3$ are fixed
by (2.15), i.e. there is no arbitrary function at this order.
\newpage
\begin{centerline}
{\bf References}
\end{centerline}
\begin{enumerate}
\item A.Ramani, B.Grammaticos and T.Bountis, Phys.Rep.180 (1989) 159.
\item M.J.Ablowitz, A.Ramani and H.Segur, J.Math.Phys.21 (1980) 715.
\item J.Weiss, M.Tabor and G.Carnevale, J.Math.Phys.24 (1983) 522.
\item J.Weiss, J.Math.Phys.24 (1983) 1405.
\item J.Weiss, J.Math.Phys.25 (1985) 2226.
\item G.Nicolis and I.Prigogine, Selforganization in nonequilibrium systems
      (Wiley Interscience. New York, 1977).
\item R.Lefever, M.Herschkowitz-Kaufman and J.W.Turner, Phys.Lett.A60 (1977)
      389.
\item C.Borzi, H.L.Frisch, R.Gianotti and J.K.Percus, J.Phys.A23 (1990) 4823.
\item P.Kaliappan, M.Lakshmanan and P.K.Ponnuswamy, J.Phys.A13 (1980) L227.
\item P.K.Vani, G.A.Ramanujam and P.Kaliappan, J.Phys.A26 (1993) L97.
\item M.Jimbo, M.D.Kruskal and T.Miwa, Phys.Lett.A92 (1982) 59.
\item S.R.Choudhury, Phys.Lett.A159 (1991) 311.
\item A.G.Webster, Partial Differential Equations of Mathematical Physics
      (Hafner Publishing Company Inc. New York, 1947), chapter IV.
\end{enumerate}
\end{document}